# Towards Models for Availability and Security Evaluation of Cloud Computing with Moving Target Defense

Matheus Torquato
*Student*
CISUC, Department of Informatics Engineering
University of Coimbra
Coimbra, Portugal
mdmelo@dei.uc.pt

Marco Vieira
*Advisor*
CISUC, Department of Informatics Engineering
University of Coimbra
Coimbra, Portugal
mvieira@dei.uc.pt

*Abstract*—Security is one of the most relevant concerns in cloud computing. With the evolution of cyber-security threats, developing innovative techniques to thwart attacks is of utmost importance. One recent method to improve cloud computing security is Moving Target Defense (MTD). MTD makes use of dynamic reconfiguration in virtualized environments to "confuse" attackers or to nullify their knowledge about the system state. However, there is still no consolidated mechanism to evaluate the trade-offs between availability and security when using MTD on cloud computing. The evaluation through measurements is complex as one needs to deal with unexpected events as failures and attacks. To overcome this challenge, we intend to propose a set of models to evaluate the availability and security of MTD in cloud computing environments. The expected results include the quantification of availability and security levels under different conditions (e.g., different software aging rates, varying workloads, different attack intensities).

## I. INTRODUCTION

Previous works show cloud computing security as a significant research challenge [1] [2]. One of the root problems for cloud security is the intrinsic advantage of attackers over defenders. Attackers can perform a series of actions (e.g., repeated attacks, vulnerability analysis) until they achieve their goal. So, the attackers can try to explore a specific system vulnerability while the defenders have to protect all the possible attack venues [3]. Besides that, the generally static nature of data centers facilitates the attacker to obtain enough information to improve the chance of attack success.

MTD is a flexible technique for system security improvement. Previous papers showed the effectiveness of MTD deployment in environments like the Internet of Things (IoT) [4], Virtualized Containers [5], Software Defined Networks (SDN) [6], and cloud computing [7]. In the cloud computing context, MTD techniques can be used to thwart or reduce the impact of security attacks as co-residency attacks [8] and Distributed Denial of Service attacks [9].

The United States Department of Homeland Security defines MTD as "*the concept of controlling change across multiple system dimensions to increase uncertainty and apparent complexity for attackers, reduce their window of opportunity and increase the costs of their probing and attack efforts.*" [10].

Besides the security concern, cloud-hosted applications also need high availability levels. There are several strategies to achieve this goal as failover techniques, redundancy, and software rejuvenation. But, the problem is to evaluate the possible availability and security impacts of applying Moving Target Defense along with availability improvement techniques.

Usually, there is a trade-off between security and availability in cloud computing systems when applying Moving Target Defense. For example, a common technique of MTD on cloud computing is Virtual Machine Migration. VM migration moves a VM from one physical machine to another. This remapping can, for example, avoid co-residency attack. However, each VM migration (even in Live Migration mode) has an associated downtime [11]. So, if we decide to perform too frequent VM migrations, we may achieve higher levels of system security, but lower availability levels. Otherwise, if the system manager decides to deploy less frequent migrations, the system may reach higher availability levels but jeopardizes the system security.

Our research aims to propose a set of models to evaluate the trade-offs between security and availability of cloud computing MTD based on different policies of VM placement. From the models' results, we will select specific policies to reach the desired levels of security and availability.

Our models will be mainly based on Stochastic Reward Nets (SRN) that are extensively used for cloud computing availability evaluation [12]. SRNs are also suitable for security evaluation [13].

The remainder of this paper is as follows. Section II presents the details of our research goals. Section III contains our research methodology. Section IV presents the related works. Section V highlights our current work. Finally, Section VI contains final remarks.

## II. GOALS

The main research question (RQ) of this work is: *What are the trade-offs between cloud computing availability and security when applying specific policies of MTD based on VM placement techniques?* This work aims to propose a set of models for availability and security evaluation of Moving Target Defense on cloud computing, aiming to answer this question.

The following RQs will drive the design of the models to be proposed:

1) *What is the availability level of cloud computing architectures?* The first step is to design a scalable model for availability evaluation of different cloud architectures.
2) *What is the security level of a given cloud deployment architecture?* The second step is to propose a model for cloud security evaluation. In this step, we will also select the considered threat models (i.e., security threats as Denial of Service, Man-in-the-middle, and Side-channel attacks). Our goal is to cover the Denial of Service (DoS) and Man-in-the-Middle attacks.
3) *What are the side-effects of MTD based on Virtual Machine (VM) placement on the system availability?* Using the model from the first step, our goal is to expand the model including MTD based on VM placement.
4) *What are the cloud computing security levels achieved by MTD techniques?* Using the model from previous steps, we aim to add the behavior related to MTD based on VM placement techniques.

Finally, we aim to merge the models obtained from the research questions mentioned above. The final model will allow evaluating the trade-offs between cloud availability and security, considering MTD based on VM placement techniques. In the future, we also intend to answer the following research questions:

1) *What are the actual levels of availability and security of MTD based on VM placement when considering aspects of software aging and rejuvenation?*
2) *What is the cloud computing performance overhead caused by MTD based on VM placement?*
3) *Is there any cloud computing reliability improvement due to MTD based on VM placement?*

## III. RESEARCH METHODOLOGY

Figure 1 presents the intended workflow and expected contributions from our research work. In the following text, we present our planned step-by-step for conducting the proposed research.

Step #1 - *Design a baseline availability model for cloud computing*. Firstly, we aim to design an availability model for general cloud computing architectures. Our previous papers [12] and [14] have the obtained results from this first step. We are now working on expanding their contribution to more complex cloud computing architectures.

Step #2 - Design a baseline security model for cloud computing. We aim to add the security aspect in our modeling framework. Our main focus in the security aspect is the *security risk* focusing on the **probability of attack success**. We consider adding realistic scenarios in the modeling process. One interesting example is proposed by the TPCx-V benchmark[1], which is based on the architecture of a brokerage firm.

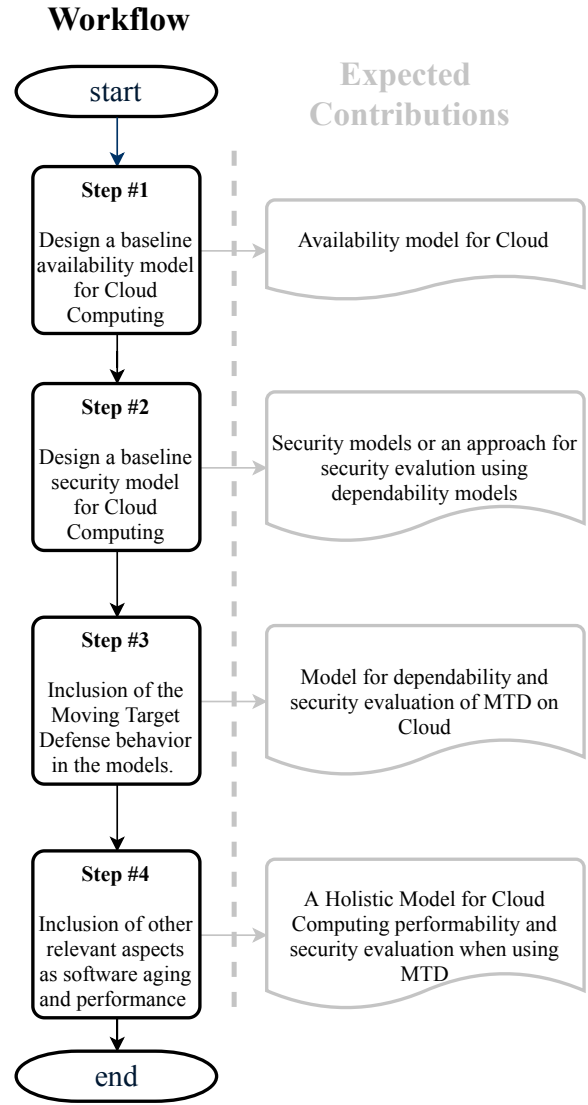

Fig. 1. Research workflow and expected contributions

Step #3 - Include MTD behavior in the models. Our goal is to cover the two main approaches of MTD deployments: 1) **MTD based on VM migration scheduling** and 2) **MTD based on creation and deletion of VMs**. Those behaviors will be incorporated into the models obtained from the previous steps. From these models, we will be able to conduct a trade-off analysis between cloud computing security and availability while using MTD based on VM placement policies.

Step #4 - Cover other relevant aspects of MTD on cloud as **software aging and rejuvenation** and **performance impacts** caused by applying MTD techniques.

---

[1] http://www.tpc.org/tpcx-v/default.asp

## IV. RELATED WORKS

Alavizadeh et al. [15] provide a comprehensive security assessment of MTD on cloud computing. The evaluation is based on modeling and analysis of MTD techniques. The authors evaluate four security metrics: system risk, attack cost, return on attack, and availability. The assessment uses on Hierarchical Attack Representation Model (HARM) models for combined MTD techniques. The main contribution of the paper is an approach to evaluate the effectiveness of combined MTD. This paper provides relevant insights for our idea of availability and security modeling of MTD deployment on cloud computing. We aim to tackle the limitations observed in the paper as: modeling the co-residency attacks, taking account of the cost of MTD on cloud computing availability and security, and including other relevant aspects on the analysis (e.g., software aging and rejuvenation, and hardware and software failure and repair).

Thebeau et al. [16] provides a theoretical point-of-view of how to measure resiliency of a cloud which applies Software Based Encryption (SBE) MTD. SBE uses software diversity to improve system security, survivability, and resilience. The paper describes some of the essential concepts of security evaluation as integrity, availability, survivability, and confidentiality. Finally, the paper proposes a model for resiliency quantification in scenarios with SBE-based MTD. We also aim to deliver models covering different MTD deployments on Cloud Computing. Such deployments are based on shuffle techniques as VM Migration and remap of the VM placement.

Ahmed and Bhargava [17] propose *Mayflies* MTD framework for distributed systems. *Mayflies* use a specific policy of VM placement as MTD. The idea is to perform VM substitution through creation and deletion cycles, obeying certain time intervals. Every cycle of substitution changes a VM characteristic. In *Mayflies*, VMs are created to use a different operating system from the previous deleted VM. The strategy avoids attack progress or the spread of an undetected attack. The authors evaluate their proposed framework through experiments in a real testbed. However, different from our intended approach, the paper does not present a security analysis of the proposed technique.

Chung et. al. [18] proposes SeReNe, a platform to deliver Network-Security-as-a-Service (NSaaS) for multi-tenant data center environments. SeReNe plans to apply MTD using diversity techniques to mitigate software vulnerabilities as Bohrbugs, Mandelbugs, and aging-related bugs. However, SeReNe is still in a conceptual phase, and the paper lacks its practical implementation and evaluation. We also intend to include the effects of MTD deployments on software aging and rejuvenation. But, different from SeReNe, our approach is focused on shuffling VM placement techniques.

The majority of the papers in the area neglects the evaluation of trade-offs between availability and security of MTD deployments on cloud computing. Different from the papers mentioned earlier, we aim to design an evaluation approach able to provide security and availability results to support the decision-making process.

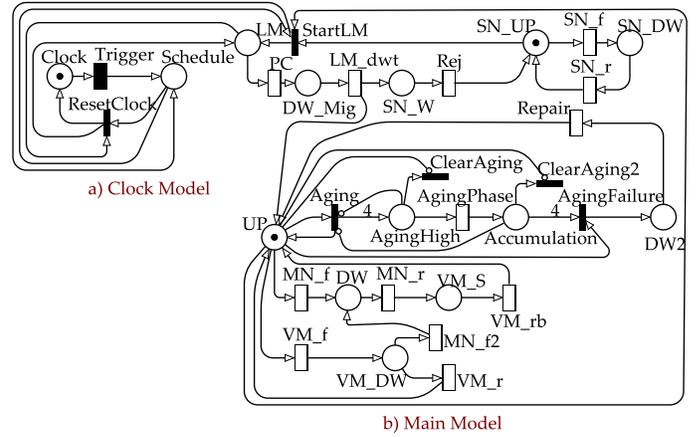

Fig. 2. Availability and Security Model

## V. CURRENT WORK

We are now working on the second step of our workflow. The papers [12] and [14] present the obtained results from the first step of our research.

Figure 2 presents the current version of our model. This model covers a simple virtualized environment with one Main Node, one Standby Node, and one Virtual Machine. This model also covers software aging and rejuvenation aspects. In this model, we consider software aging effects in the Virtual Machine Monitor (VMM) software component [19]. VMM software rejuvenation is supported by VM migration scheduling.

Specifically in the model, the places `Clock` and `Schedule` in conjunction with the transitions `Trigger` and `ResetClock` represent the behavior of a system clock for VM migration submission in the environment. The places `LM`, `DW_Mig` and `SN_W` with the transitions `StartLM`, `PC`, `LM_dwt` and `Rej` represent the behavior of the VM migration. The Erlang sub-net[2] represent the software aging accumulation process. The remainder places and transitions are related with the failure and repair behavior of the architectural components.

The model consists of an SRN model. In the SRN models, we can represent the failure and repair behavior for each component of the virtualized environment. As the SRN models are already extensively used for availability evaluation, one of the major challenges is to extract security measures from the availability model.

So, besides extracting availability measures, we also compute a security measure named RISKSCORE[3] from the proposed availability model. Figure 3 presents the obtained results. Our obtained results include system unavailability and the RISKSCORE related to Man-in-the-Middle (MITM) and Denial of Service (DoS) threats. MITM attacks have high severity and can harm system confidentiality. DoS attacks are among the most relevant threats for cloud computing high availability. The X-Axis represents the considered software rejuvenation policies. `Rej. Trigger` means the adopted time interval between VM migrations.

[2]Places `AgingHigh`, `Accumulation` and `DW2` with the transitions `Aging`, `AgingPhase`, `AgingFailure` and `ClearAging(2)`
[3]RISKSCORE is obtained through the steady-state probability of the system being in a risky state (from a security perspective).

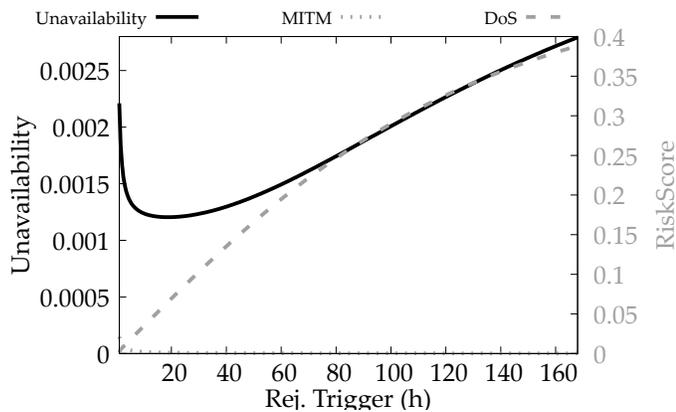

Fig. 3. Obtained results

From the obtained results it is possible to notice that there is a specific VM migration policy which minimizes system unavailability. Besides that, there are other VM migration policies which minimize the RISKSCORE related to Denial of Service or Man-In-The-Middle attacks. Therefore, the VM migration policy selection will depend on the assigned weight for the considered metrics.

We aim to expand the model in two ways: i) adopting larger architectures with more nodes and VMs; ii) to cover more realistic scenarios like what is presented in the TPC-xV architecture and iii) to consider other security threats.

## VI. FINAL REMARKS

This paper presented the work-in-progress of our research. This research tackles the cloud security problem from a different perspective, aiming not only on security improvement but also in the possible availability impact due to security mechanisms adoption. In the context of this research, we focus only on the Moving Target Defense technique.

Our research is on the initial stage, and we are working to improve our proposed model to more realistic scenarios. However, our research effort so far produced results published in two research papers [12] and [14].

Our next step is to proceed with the submission of the model's current version and results to validate our security evaluation approach. We aim to deal with the state-explosion problem using interacting models approach.

The expected contributions of our research will advance the state of the art, providing a holistic model for performability and security evaluation of a cloud computing environment with Moving Target Defense.